\documentclass[12pt,superscriptaddress,aps,prd,preprint,showpacs]{revtex4}
\usepackage{graphicx}
\usepackage{amsfonts}
\usepackage{slashed}
\usepackage[latin1]{inputenc}
\usepackage{amsmath}
\usepackage{hyperref}
\newcommand{\bea}{\begin{eqnarray}}
\newcommand{\eea}{\end{eqnarray}}

\begin{document}

\title{Nonrelativistic effective potential of the bumblebee model}

\author{C. Furtado}
\affiliation{Departamento de F\'{\i}sica, Universidade Federal da Para\'{\i}ba,\\
 Caixa Postal 5008, 58051-970, Jo\~ao Pessoa, Para\'{\i}ba, Brazil}
\email{furtado,jroberto,petrov,pporfirio@fisica.ufpb.br}

\author{J. R. Nascimento}
\affiliation{Departamento de F\'{\i}sica, Universidade Federal da Para\'{\i}ba,\\
 Caixa Postal 5008, 58051-970, Jo\~ao Pessoa, Para\'{\i}ba, Brazil}
\email{furtado,jroberto,petrov,pporfirio@fisica.ufpb.br}

\author{A. Yu. Petrov}
\affiliation{Departamento de F\'{\i}sica, Universidade Federal da Para\'{\i}ba,\\
 Caixa Postal 5008, 58051-970, Jo\~ao Pessoa, Para\'{\i}ba, Brazil}
\email{jfurtado,roberto,petrov,pporfirio@fisica.ufpb.br}

\author{P. Porfirio}
\affiliation{Departamento de F\'{\i}sica, Universidade Federal da Para\'{\i}ba,\\
 Caixa Postal 5008, 58051-970, Jo\~ao Pessoa, Para\'{\i}ba, Brazil}
\email{furtado,jroberto,petrov,pporfirio@fisica.ufpb.br}

\begin{abstract}
In this paper, we explicitly obtain the nonrelativistic Breit potential in the bumblebee model arising in the weak gravity limit of the metric-affine bumblebee gravity, coupled to the spinor matter. In this theory,  in the lower (second) order in the small coupling constant $\xi$ (and the second order in the Lorentz-violating (LV) vector $\beta_{\mu}$) it demonstrates the $1/r$ asymptotics, which naturally corresponds to the massless character of the theory, while higher orders in $\xi$ yield anisotropic modifications of the Coulomb potential due to the Lorentz symmetry breaking. For the lower-order modification of the effective potential, we calculate LV corrections to energy levels of the hydrogen atom.
\end{abstract}


\maketitle

\section{Introduction}

The spontaneous symmetry breaking is actually considered as the most appropriate mechanism for implementing the Lorentz symmetry violation (LSV) \cite{KosGra,KosLiGra}. The typical model considered within this framework is the bumblebee model proposed in \cite{KosGra}, calling special attention since it provides the most natural manner to implement LV scenarios into a curved space-time. Different studies related with this model, including its generalization for the curved space-time, with subsequent analysis of various related issues, exact solutions, see e.g. \cite{Bertolami:2005bh,Capelo:2015ipa,Casana,Ovgun:2018xys,Li:2020dln,Maluf:2020kgf,Gullu:2020qzu,Xu:2026zgd,Liu:2025oho,Zhu:2026vae} weak-field corrections \cite{Bailey:2006fd,Bluhm:2007bd}, modifications of dispersion relations and loop calculations for the weak gravity limit (see e.g. \cite{Maluf:2014dpa,Belchior:2023cbl}) 
have been carried out.

The interest to the model, together with the idea that the spontaneous LSV (as well as, in principle, spontaneous breaking of any symmetry) is the essentially low-energy phenomenon, naturally calls attention to studies of nonrelativistic limit of the bumblebee model, in particular, of the nonrelativistic potential (that is, the Breit potential) in this model, which can naturally arise within studies of particle scattering. Earlier studies of the nonrelativistic potential in LV theories have been performed for the intermediating scalar field \cite{Altschul,aether} and for the gauge one in \cite{Ferreira:2004hx} (it is worth to mention also studies of tree-level scattering in LV theories carried out in \cite{Casana:2012vu,Charneski:2012py,Toniolo:2016oom}), therefore it is natural to consider this effective potential for the bumblebee field. Differently from the previous studies, we find the nonrelativistic effective potential for the weak-field regime of the metric-affine bumblebee gravity, which is novel in itself. Furthermore, the effective dynamics of the model is described in terms of the vector field, which acts as the intermediate field, while the spinor-vector vertices are essentially nonminimal. Within our study, we follow the methodology developed for three-dimensional theories \cite{Girotti1,Girotti2} to obtain the tree-level scattering potential and generalize it to the four-dimensional space-time, for the case when the intermediate particle is the bumblebee vector boson defined within the effective bumblebee model studied in \cite{Delhom:2020gfv}. It should be noted that this model arises in the weak gravity limit of the metric-affine bumblebee gravity formulated originally in \cite{Delhom:2019wcm}.

The structure of our paper looks as follows. In Section 2, we review our bumblebee theory coupled to spinors and write down its propagator. In Section 3, we calculate the Breit potential. The results are discussed in Section 4.

\section{Coupling of spinors to the bumblebee field}

To study the spinor-spinor scattering within our model, we start with the Lagrangian of the effective bumblebee theory coupled to spinors $\Psi$ defined in \cite{Delhom:2020gfv}:
\begin{eqnarray}
	\nonumber\label{ap2}\mathcal{L}_{spEF}&=&\bar{\Psi}\left(i\gamma^{\mu}\partial_{\mu}-m\right)\Psi-\frac{i}{4}\xi\bar{\Psi}B^{\nu}B_{\nu}\gamma^{\mu}\partial_{\mu}\Psi-
	\frac{i}{2}\xi\bar{\Psi}\left(B^{\nu}\gamma_{\nu}\right)B^{\mu}\partial_{\mu}\Psi+\frac{\xi}{2} m B^{\nu}B_{\nu} \bar{\Psi}\Psi-\\
	\nonumber&-&i\frac{\xi}{4}\bar{\Psi}\Big(B_{\alpha}\left(\partial_{\mu} B^{\alpha}\right)+B^{\nu}\left(\partial_{\nu} B_{\mu}\right)+
	(\partial_\alpha B^\alpha) B_{\mu}\Big)\gamma^{\mu}\Psi+\mathcal{O}(\xi^2),\label{SpinLagPert}\\
	\nonumber{\cal L}_{BEF}&=&-\frac{1}{4}B_{\mu\nu}B^{\mu\nu}+\frac{M^2}{2}B^2-\frac{\Lambda}{4}(B^2)^2+\\ 
	&+&\frac{\xi}{2}\Big[B^{\mu\nu}B^\alpha{}_{\nu}B_\mu B_\alpha-\frac{1}{4}B_{\mu\nu}B^{\mu\nu}B^2-\frac{3}{4}\Lambda(B^2)^3\Big]+
	\mathcal{O}(\xi^2).
	\label{BumbLagPert}
\end{eqnarray} 
This theory represents itself as the effective model describing the weak gravity limit of the metric-affine bumblebee gravity coupled to a spinor matter (see details in \cite{Delhom:2020gfv}). Here $B_{\mu}$ is the bumblebee field characterized by the potential $V=-\frac{M^2}{2}B^2+\Lambda\frac{(B^2)^2}{4}+O(\xi)$, possessing an infinite set of minima, and $\xi$ is the nonmimimal bumblebee-gravity coupling constant assumed to be small, so, $|\xi\beta^2|\ll 1$.

To introduce the spontaneous Lorentz symmetry breaking, we can define the vacuum $\beta_{\mu}$ satisfying the relation $\beta^2=\frac{2M}{\Lambda}+O(\xi)$ for $\xi\neq 0$ and $\beta^2=\frac{M}{\Lambda}$ for $\xi=0$. Hence we see that $\beta^{\mu}$ is time-like. As a result, we can shift $B_{\mu}\to\tilde{B}_{\mu}+\beta_{\mu}$, where $\tilde{B}_{\mu}$ is the bumblebee fluctuation around the new vacuum $\beta_{\mu}$, and write down, first, the action coupling the spinor field to the bumblebee one in the presence of the LV background $\beta_{\mu}$ (cf. \cite{Delhom:2020gfv}),
\begin{eqnarray}
\nonumber{\cal L}_{sp}&=&\bar{\Psi}\bigg[i\gamma^{\mu}\partial_{\mu}-m\left(1-\frac{1}{2}\xi\beta^{2}\right)-
\frac{i}{2}\xi\left(\beta^{\mu}\beta^{\nu}+\frac{1}{2}\beta^2\eta^{\mu\nu}\right)\gamma_{\mu}\partial_{\nu}\bigg]\Psi-\nonumber\\
\nonumber&-&\bar{\Psi}\bigg[\frac{i}{2}\xi\big(\tilde{B}^{\mu}\beta^{\nu}+\tilde{B}^{\nu}\beta^{\mu}+
(\tilde{B}\cdot\beta)\eta^{\mu\nu}\big)\gamma_{\mu}\partial_{\nu}-\xi m\tilde{B}^{\nu}\beta_{\nu}\bigg]\Psi-\nonumber\\
&-&\bar{\Psi}\left[\frac{i}{2}\xi\left(\tilde{B}^{\mu}\tilde{B}^{\nu}+\frac{1}{2}\tilde{B}^2\eta^{\mu\nu}\right)\gamma_{\mu}\partial_{\nu}-\frac{1}{2}\xi m\tilde{B}^2\right]\Psi-\nonumber\\
\nonumber&-&\frac{i}{4}\xi\bar{\Psi}\bigg[\tilde{B}_{\alpha}(\partial_{\mu} \tilde{B}^{\alpha})+\beta_{\alpha}(\partial_{\mu} \tilde{B}^{\alpha})+\tilde{B}^{\nu}(\partial_{\nu} \tilde{B}_{\mu})+\\
&+&\beta^{\nu}(\partial_{\nu} \tilde{B}_{\mu})+(\partial_\alpha \tilde{B}^\alpha) \tilde{B}_{\mu}+(\partial_\alpha \tilde{B}^\alpha) \beta_{\mu} \bigg]\gamma^{\mu}\Psi,\label{SpinLagVEV}
\end{eqnarray}
second, the complete propagator of the bumblebee field which is only interesting for us within our study of the spinor-spinor scattering \cite{Delhom:2020gfv}:
\bea
\label{complete}
<\tilde{B}_{\alpha}(-k)\tilde{B}_{\beta}(k)>&=&G_{\alpha\beta}=i\frac{1}{-k^2(1+\frac{\xi\beta^2}{2})+\xi(\beta\cdot k)^2}\Big[\eta_{\alpha\beta}-
\\\nonumber &-&
\Delta^{-1}\Big([-k^2(1-\frac{1}{2}\xi\beta^2)+\xi(\beta\cdot k)^2-\nonumber\\
&-&2\Lambda\beta^2](1+\frac{\xi\beta^2}{2})k_{\alpha}k_{\beta}+
\xi(\beta\cdot k)^2(-2\Lambda+\xi k^2)\beta_{\alpha}\beta_{\beta}-\nonumber\\\nonumber&-&
(1+\frac{\xi\beta^2}{2})
(-2\Lambda+\xi k^2)(\beta\cdot k)(\beta_{\alpha}k_{\beta}+\beta_{\beta}k_{\alpha})
\Big)\Big],
\eea
where
\begin{eqnarray}
\Delta&=&-(-2\Lambda+\xi k^2)(\beta\cdot k)^2(1+\frac{\xi\beta^2}{2})+\nonumber\\&+&
\xi(\beta\cdot k)^2[-k^2(1-\frac{1}{2}\xi\beta^2)-2\Lambda\beta^2+\xi(\beta\cdot k)^2].
\end{eqnarray}
We note that, as $\xi$ is small, one can use the following simplified form of the propagator:
\bea
\label{simple}
G^{(0)}_{\alpha\beta}(k)&=&-i\frac{1}{k^2}\Big[\eta_{\alpha\beta}-
\frac{1}{2\Lambda(\beta\cdot k)^2}\Big([-k^2-2\Lambda\beta^2]k_{\alpha}k_{\beta}+\\
\nonumber&+&2\Lambda(\beta\cdot k)(\beta_{\alpha}k_{\beta}+\beta_{\beta}k_{\alpha})
\Big)\Big]
+O(\xi).
\eea
The full-fledged propagator (\ref{complete}), as well as its simplified form (\ref{simple}) can be used to obtain the effective potential.

Then, it follows from the action (\ref{ap2}) that all spinor-vector vertices are proportional to at least the first order in $\xi$, and we note that there is no "minimal" coupling $\bar{\Psi}\gamma^{\mu}B_{\mu}\Psi$, analogous to that on present in QED,  in our theory. As a result, the standard diagram describing the spinor-spinor scattering will be proportional to at least $\xi^2$, hence, to obtain the corresponding contribution in the leading order it is sufficient to calculate the vector propagator in the zero order in $\xi$, and to use the external spinor field satisfying the standard, uncorrected Dirac equation and hence independent on the coupling $\xi$.

\section{Breit potential}

As it is well known (see e.g. \cite{Das}), in the nonrelativistic limit the dominant term of $\bar{\psi}\gamma^{\mu}\psi$ is that one corresponding to $\mu=0$ (i.e. the static potential dominates), i.e. $\bar{\psi}\gamma^0\psi=\phi^*\phi$ (where the Dirac spinor $\Psi$ is described by two Weyl spinors $\phi$ and $\chi$, as usual, see e.g. \cite{Das}), so, effectively we need only $G_{00}(k)$, i.e. the bumblebee field can be chosen as $B^{\mu}=(\Phi,\vec{0})$, which is a natural analogue of the static limit of the electrodynamics.

Then, the transferred momentum is $k=p-p^{\prime}$, in the nonrelativistic limit $p=(m,\vec{p})$ and $p^{\prime}=(m,\vec{p}^{\prime})$, so, $k=(0,\vec{p}-\vec{p}^{\prime})=(0,\vec{k})$, and $k^2=-\vec{k}^2$ (this is rather standard situation for the nonrelativistic limit, see e.g. \cite{aether}). As we consider the $B^{\mu}$ along the time axis, to consider the static limit, it is natural to require $\beta^{\mu}$ also along the same axis, to justify the fact that $\beta^{\mu}$ is the v.e.v. of $B^\mu$. However, in this case we face a difficulty. Indeed, the simple choice for the time-like $\beta^{\mu}$, i.e. $\beta^{\mu}=(\beta,0)$ implies $(\beta\cdot k)=0$, so, our propagator, both simplified (\ref{simple}) and complete one (\ref{complete}), formally becomes strongly singular. Thus, the nonrelativistic limit needs a more careful treatment since  the straightforward use of this approximation naively seems to be impossible. So, let us, for a time being, choose $\beta^{\mu}=(\beta_0,\vec{\beta})$ so that $\beta_0^2-\vec{\beta}^2=\beta^2$, to guarantee $\beta\cdot k=\vec{\beta}\cdot\vec{k}\neq 0$, with $|\beta_0|\gg|\vec{\beta}|$, to ensure the $\beta^{\mu}$ to be time-like.  At the same time, for such a choice of $\beta^{\mu}$, taking  into account that $k_{\mu}=(0,\vec{k})$, one sees that, as $k_0=0$, one has $G_{00}=-i/k^2+O(\xi)$ in (\ref{simple}), while $\vec{\beta}\neq 0$, and, by continuity reasons, one can complete this component of the Green function by requirement that $G_{00}=-i/k^2+O(\xi)$ while $\vec{\beta}=0$ as well, i.e. actually the singularity of the $G_{00}$ component of the propagator (\ref{simple}) at $\vec{\beta}=0$ turns out to be removable. Finally,
 one has (here and further we multiply this expression by $-i$, to match definitions within electrodynamics):
\bea
\label{Green0}
G^{(0)}_{00}(k)&=&<\Phi(\vec{k})\Phi(-\vec{k})>=\frac{1}{\vec{k}^2}
+O(\xi).
\eea
This Green function is $\beta$-independent and free of any singularities at $\vec{\beta}=0$, so, we can choose $\beta^{\mu}=(\beta,0)$ henceforth without any difficulties. We note also that this propagator is massless which matches the fact that the bumblebee theory near the minima behaves as the massless one, see the discussion in \cite{bumbvect}.

The solution of the Dirac equation relevant in the nonrelativistic limit corresponds to the positive energy (as we consider spinor-spinor scattering) and looks like (see e.g. \cite{Das}):
\bea
\label{psi}
\Psi(\vec{r},t)=C\left(\begin{array}{c}
\phi_0(\vec{r})\\\frac{\vec{\sigma}\cdot\vec{p}}{2m}\phi_0(\vec{r})
\end{array}
\right)e^{-iEt},
\eea
with $C$ is a normalization constant (we note again that we choose it to be equal to 1: indeed, the standard choice for it is $C=\sqrt{\frac{E+m}{2m}}$, so, at $E\simeq m$, as it must be in the nonrelativistic limit, one has $C=1$), 
and $\phi_0(\vec{r})$ is the two-component spinor: as our example, we choose the simplest case $\phi_0=\left(\begin{array}{c} 0\\
1
\end{array}
\right)e^{i\vec{p}\cdot\vec{r}}$, while for the spinor $\left(\begin{array}{c} 1\\
0
\end{array}
\right)e^{i\vec{k}\cdot\vec{r}}$ the calculations are analogous.
Then, as the tree-level scattering amplitude of spinors is given by $\Gamma=(\bar{\Psi}\Gamma^{\mu}\Psi)G_{\mu\nu}(\bar{\Psi}\Gamma^{\nu}\Psi)$ (cf. \cite{Altschul}),  with the component $G_{00}$ of the Green function dominates in the nonrelativistic limit, and $\Gamma^{\mu}$ is the matrix describing the vertex $\bar{\Psi}\Gamma^{\mu}\Psi B_{\mu}$ defined by (\ref{effvertex}).
Explicitly, we can write our Dirac spinor $\Psi$ (\ref{psi}) in the nonrelativistic limit as 
(cf. e.g. \cite{Das}):
\begin{eqnarray}
\label{nonrelspin}
\Psi(\vec{r},t)=Ce^{-iEt+i\vec{p}\cdot\vec{r}}\left(\begin{array}{c}
0\\ 1\\ \frac{p_z}{E+m}\\\frac{p_x+ip_y}{E+m}
\end{array}
\right),
\end{eqnarray}

 Thus, the relevant corrections to the effective potential arise only from the new $\xi$-dependent vertices. We note that there is no minimal (standard) vector-spinor coupling like $\bar{\Psi}\gamma ^{\mu}B_{\mu}\Psi$ in our theory, hence our calculation will be strongly different from the usual QED case. Actually, we must calculate the vertex
 \bea
 \label{effvertex}
 V&=&-\xi\bar{\Psi}\bigg[\frac{i}{2}\big(\tilde{B}^{\mu}\beta^{\nu}+\tilde{B}^{\nu}\beta^{\mu}+
(\tilde{B}\cdot\beta)\eta^{\mu\nu}\big)\gamma_{\mu}\partial_{\nu}-m\tilde{B}^{\nu}\beta_{\nu}\bigg]\Psi-\nonumber\\
&-&\frac{i}{4}\xi\bar{\Psi}\bigg[\beta_{\alpha}(\partial_{\mu} \tilde{B}^{\alpha})+
\beta^{\nu}(\partial_{\nu} \tilde{B}_{\mu})+(\partial_\alpha \tilde{B}^\alpha) \beta_{\mu} \bigg]\gamma^{\mu}\Psi,\label{vertex}
 \eea
 since quartic vertices do not contribute to the scattering amplitude at the tree level (indeed, they are essential only for the loop corrections).  As we already noted, we consider the case where the bumblebee field $B^{\mu}$ is directed along the time axis as well as its v.e.v. $\beta^{\mu}$, i.e. $\tilde{B}^{\mu}=(\Phi,0,0,0)$ and $\beta^{\mu}=(\beta,0,0,0)$. Substituting these values to the vertex (\ref{effvertex}), we rewrite it as:
 \bea
  V&=&-\xi\bar{\Psi}\bigg[\frac{i}{2}\big(2\beta\Phi\gamma^0\partial_0 +\beta\Phi\gamma^{\mu}\partial_{\mu})- m\beta\Phi+
\frac{i}{4}(
\beta(\partial_{\mu}\Phi)\gamma^{\mu}+2\beta(\partial_0\Phi)\gamma^0)\bigg]\Psi.
\label{vertex1}
 \eea
It remains to use here the nonrelativistic limit of our spinors (\ref{nonrelspin}). It is easy to check that $\bar{\Psi}\gamma^i\Psi\simeq C^2\dfrac{p^i}{m}$, and, as $|\vec{p}/m|\ll 1$ in our limit, we see that the term $\bar{\Psi}\gamma^i\Psi$ is suppressed in comparison with $\bar{\Psi}\gamma^0\Psi\simeq C^2$ in the nonrelativistic regime. Similarly, the $\bar{\Psi}\gamma^i\partial_i\Psi$ is suppressed in comparison with $\bar{\Psi}\gamma^0\partial_0\Psi$ in the same regime. Further, the field $\Phi(\vec{x})$ is assumed to be static as we consider the Breit potential, and the zero component of the four-momentum $k_{\mu}$ playing the role of the argument of the Green function $G_{00}$, is zero in the static case, hence we can assume $\partial_0\Phi\simeq 0$. 
Therefore we can simplify our vertex as
 \bea
  V&=&-\beta\xi\bar{\Psi}\bigg[\frac{3i}{2}\Phi\gamma^0\partial_0- m\Phi+
\frac{i}{4}
(\partial_i\Phi)\gamma^i\bigg]\Psi.
\label{vertex2}
 \eea
 Now, we use $i(\bar{\Psi}\gamma^0\partial_0\Psi)\simeq C^2E\simeq C^2m$, $\bar{\Psi}\Psi=C^2(1+O(\frac{\vec{p}^2}{m^2}))$, $\bar{\Psi}\gamma^i\Psi\simeq\frac{p^i}{m}C^2$. Then, as we noted above, in the nonrelativistic limit we choose $C=1$; also, in the static limit we can neglect $p_i$ in comparison with $m$ as well and assume $|\partial_i\Phi|\ll m|\Phi|$. So, our scattering amplitude formed by the contraction of $\Phi$ fields from two vertices $V$ into the Green function (\ref{Green0}) taken in the static limit where the $G_{00}$ component, corresponding to the static fields, dominates, after the inverse Fourier transform, takes the form 
 \bea
 \Gamma=-\frac{m^2(\beta\xi)^2}{16\pi r}.
 \eea
 This is the additive correction to the Coulomb potential which we assumed to be the usual one, $V_C=-\frac{1}{4\pi r}$. Although this correction is generated by the Lorentz symmetry breaking it is spherically symmetric which is consistent with the fact that our LV vector is directed along the time axis and does not introduce the spatial anisotropy. We can find the corrections to the energy levels in the hydrogen atom caused by this modification of the Coulomb potential (we note that the LV correction to the potential has the same sign as the classical potential itself, hence the LV effect increases the "electrostatic" attraction of particles). Thus, we face the problem to calculate the corrections to the energy levels for the effective potential
\bea
\label{veff}
V_{eff}=-\frac{1}{4\pi r}(1+(m\beta\xi/2)^2).
\eea
We note that in this expression, the $m$ is the mass of the fermions, it cannot be equal to zero since in this case the nonrelativistic limit is senseless.

As our corrected potential continues to be the central one within our approximation, the energy levels can be found through redefinition of constants. Indeed, as it is well known, for the hydrogen-like atom with the potential $V=-\frac{\zeta}{4\pi r}$, the energy levels are $E_n=-\frac{m\zeta^2}{2n^2}$. In our case, $\zeta=1+(m\beta\xi/2)^2$, see (\ref{veff}). So, we can find the shift of an arbitrary $n$-th energy level as $\delta E_n=E_n-E_n|_{\zeta=0}=-\frac{m}{n^2}(m\beta\xi/2)^2+O(\xi^4)$. The wave functions are modified accordingly. We note that this result is exact, without quantum mechanical perturbative corrections. 

However, such corrections will arise if we consider the higher orders in $\xi$. For example, the common denominator $-k^2(1+\frac{\xi\beta^2}{2})+\xi(\beta\cdot k)^2$ of the propagator (\ref{complete}) will yield an anisotropic potential while $\vec{\beta}\neq 0$ (we note that this denominator is very similar to that one arising in \cite{aether}, where the resulting effective potential displayed anisotropic behavior), which will imply in the nontrivial first perturbative correction to the energy levels which however will be proportional to $\xi^3$ and hence highly suppressed.

To close the discussion, let us do some numerical estimations. The absolute value of the shift of the first energy level is $|\delta E_1|=m^3\beta^2\xi^2/4$. 
Assuming $\xi\beta^2\simeq 4.9\cdot 10^{-12}$ (cf. f.e. \cite{Filho:2022yrk}), and $m=0.5$ MeV, that is, the electron mass, and assuming that the energy shift is compatible with the hyperfine energy shift equal to $5.9\cdot 10^{-6}$ eV for the hydrogen atom \cite{Kuzmak} to provide that the LV effect does not exceed the hyperfine energy shift, we can estimate the $\xi$ parameter as $10^{-11}$ ${\rm eV}^{-2}$, which implies $\beta\simeq$ 1 eV, which is an inadequately high value. Therefore, we conclude that the energy shift due to the LV effects must be much smaller than even the hyperfine splitting. Requiring, otherwise, $\beta\simeq 10^{-16}$ GeV, which is one of the largest estimations for the LV vector from \cite{datatables}, we find $|\delta E_1|\simeq 0.75\cdot 10^{-28}$ eV, that is, the value is extremely tiny in comparison even with the hyperfine splitting. It is worth noting that such energy scales cannot be probed with current detectors. However, next-generation detectors may be able to probe these scales in the foreseeable future, thereby allowing the direct observation of signatures of LSV.

\section{Summary}

Let us discuss our results. We considered the nonrelativistic limit of the bumblebee model and calculated the Breit potential in this limit.  In the lower (second) order in $\xi$, the Breit effective potential behaves as $\frac{1}{r}$, similarly to the electrodynamics. This result is consistent with the observation made in our earlier papers \cite{Delhom:2020gfv,bumbvect} that the bumblebee theory behaves as the massless theory which is consistent with the Goldstone theorem since we study our bumblebee model near its vacuum. However, in third and higher orders in $\xi$ the Breit potential is anisotropic which reflects the LV effects, similarly to the case of the scalar aether theory \cite{aether}. Moreover, in higher orders in $\xi$ the external spinors should satisfy the modified Dirac equation and hence also will be corrected.  

Also, we obtained the lower corrections to the energy levels of the hydrogen-type atom which turns out to be proportional to $(\xi\beta)^2$, so, they are also of the second order in $\xi$ (it is worth to mention that the lower one-loop correction in the spinor sector is also of the second order in $\xi$ \cite{Delhom:2020gfv}). However, it is important to mention that such corrections, although they allow to estimate either the value of $\xi$ or the LV vector $\beta^{\mu}$, are extremely tiny. In principle, it is possible to calculate the perturbative corrections of third and higher orders in $\xi$, both to energy levels and to wave functions of the hydrogen-like atom. We hope to perform such studies in one of our next papers.
 

{\bf Acknowledgments.}  This work was partially supported by Conselho Nacional de Desenvolvimento Cient\'\i fico e Tecnol\'ogico (CNPq). The work of A. Yu.\ P. has been partially supported by the CNPq project No. 303777/2023-0.


\begin{thebibliography}{100}

\bibitem{KosGra} V.~A.~Kostelecky,
Phys. Rev. D \textbf{69} (2004), 105009
[arXiv:hep-th/0312310 [hep-th]].

\bibitem{KosLiGra} V.~A.~Kosteleck\'y and Z.~Li,
Phys. Rev. D \textbf{103} (2021) no.2, 024059
[arXiv:2008.12206 [gr-qc]].

\bibitem{Bertolami:2005bh}
O.~Bertolami and J.~Paramos,
Phys. Rev. D \textbf{72} (2005), 044001
[arXiv:hep-th/0504215 [hep-th]].

\bibitem{Capelo:2015ipa}
D.~Capelo and J.~P{\'a}ramos,
Phys. Rev. D \textbf{91} (2015) no.10, 104007
[arXiv:1501.07685 [gr-qc]].

\bibitem{Casana}
R.~Casana, A.~Cavalcante, F.~P.~Poulis and E.~B.~Santos,
Phys. Rev. D \textbf{97} (2018) no.10, 104001
[arXiv:1711.02273 [gr-qc]].

\bibitem{Ovgun:2018xys}
A.~{\"O}vg{\"u}n, K.~Jusufi and {\.I}.~Sakall{\i},
Phys. Rev. D \textbf{99} (2019) no.2, 024042
[arXiv:1804.09911 [gr-qc]].

\bibitem{Li:2020dln}
Z.~Li and A.~{\"O}vg{\"u}n,
Phys. Rev. D \textbf{101} (2020) no.2, 024040
[arXiv:2001.02074 [gr-qc]].

\bibitem{Maluf:2020kgf}
R.~V.~Maluf and J.~C.~S.~Neves,
Phys. Rev. D \textbf{103} (2021) no.4, 044002
[arXiv:2011.12841 [gr-qc]].

\bibitem{Gullu:2020qzu}
{\.I}.~G{\"u}ll{\"u} and A.~{\"O}vg{\"u}n,
Annals Phys. \textbf{436} (2022), 168721
[arXiv:2012.02611 [gr-qc]].

\bibitem{Xu:2026zgd}
R.~Xu, Z.~F.~Mai and D.~Liang,
Phys. Lett. B \textbf{875} (2026), 140364
[arXiv:2601.18809 [gr-qc]].

\bibitem{Liu:2025oho}
J.~Z.~Liu, S.~P.~Wu, S.~W.~Wei and Y.~X.~Liu,
``Exact Black Hole Solutions in Bumblebee Gravity with Lightlike or Spacelike VEVS,''
[arXiv:2510.16731 [gr-qc]].

\bibitem{Zhu:2026vae}
J.~Zhu and H.~Li,
``New Exact Vacuum Solutions in Extended Bumblebee Gravity,''
[arXiv:2604.09464 [gr-qc]].

\bibitem{Bailey:2006fd}
Q.~G.~Bailey and V.~A.~Kostelecky,
Phys. Rev. D \textbf{74} (2006), 045001
[arXiv:gr-qc/0603030 [gr-qc]].

\bibitem{Bluhm:2007bd}
R.~Bluhm, S.~H.~Fung and V.~A.~Kostelecky,
Phys. Rev. D \textbf{77} (2008), 065020
[arXiv:0712.4119 [hep-th]].

\bibitem{Maluf:2014dpa}
R.~V.~Maluf, C.~A.~S.~Almeida, R.~Casana and M.~M.~Ferreira, Jr.,
Phys. Rev. D \textbf{90} (2014) no.2, 025007
[arXiv:1402.3554 [hep-th]].

\bibitem{Belchior:2023cbl}
F.~M.~Belchior and R.~V.~Maluf,
Phys. Lett. B \textbf{844} (2023), 138107
[arXiv:2307.14252 [hep-th]].

\bibitem{Altschul}
B.~Altschul,
Phys. Lett. B \textbf{639} (2006), 679-683
[arXiv:hep-th/0605044 [hep-th]].

\bibitem{aether}
M.~Gomes, J.~R.~Nascimento, A.~Y.~Petrov and A.~J.~da Silva,
Phys. Rev. D \textbf{81} (2010), 045018
[arXiv:0911.3548 [hep-th]].

\bibitem{Ferreira:2004hx}
M.~M.~Ferreira, Jr.,
Phys. Rev. D \textbf{70} (2004), 045013
[arXiv:hep-th/0403276 [hep-th]].


\bibitem{Casana:2012vu}
R.~Casana, M.~M.~Ferreira, R.~V.~Maluf and F.~E.~P.~dos Santos,
Phys. Rev. D \textbf{86} (2012), 125033
[arXiv:1212.6230 [hep-th]].

\bibitem{Charneski:2012py}
B.~Charneski, M.~Gomes, R.~V.~Maluf and A.~J.~da Silva,
Phys. Rev. D \textbf{86} (2012), 045003
[arXiv:1204.0755 [hep-ph]].

\bibitem{Toniolo:2016oom}
G.~R.~Toniolo, H.~G.~Fargnoli, L.~C.~T.~Brito and A.~P.~B.~Scarpelli,
Eur. Phys. J. C \textbf{77} (2017) no.2, 108
[arXiv:1603.01738 [hep-th]].

\bibitem{Girotti1} H.~O.~Girotti, M.~Gomes and A.~J.~da Silva,
Phys. Lett. B \textbf{274} (1992), 357-362.

\bibitem{Girotti2} H.~O.~Girotti, M.~Gomes, J.~L.~deLyra, R.~S.~Mendes, J.~R.~S.~Nascimento and A.~J.~da Silva,
Phys. Rev. Lett. \textbf{69} (1992), 2623-2626
[arXiv:hep-th/9210158 [hep-th]].

\bibitem{Delhom:2020gfv}
A.~Delhom, J.~R.~Nascimento, G.~J.~Olmo, A.~Y.~Petrov and P.~J.~Porf{\'\i}rio,
Phys. Lett. B \textbf{826} (2022), 136932
[arXiv:2010.06391 [hep-th]].

\bibitem{Delhom:2019wcm}
A.~Delhom, J.~R.~Nascimento, G.~J.~Olmo, A.~Y.~Petrov and P.~J.~Porf{\'\i}rio,
Eur. Phys. J. C \textbf{81} (2021) no.4, 287
[arXiv:1911.11605 [hep-th]].

\bibitem{Das} A. Das, Lectures on quantum field theory, World Scientific, 2008.


\bibitem{bumbvect} A.~C.~Lehum, J.~R.~Nascimento, A.~Y.~Petrov and P.~J.~Porfirio,
Gen. Rel. Grav. \textbf{57} (2025) no.66, 66
[arXiv:2402.17605 [hep-th]].

\bibitem{Kuzmak} A.~R.~Kuzmak,
J. Phys. Stud. \textbf{28} (2024) no.3, 3901.

\bibitem{datatables} V.~A.~Kostelecky and N.~Russell,
Rev. Mod. Phys. \textbf{83} (2011), 11-31
[arXiv:0801.0287 [hep-ph]].

\bibitem{Filho:2022yrk}
A.~A.~A.~Filho, J.~R.~Nascimento, A.~Y.~Petrov and P.~J.~Porf{\'\i}rio,
Phys. Rev. D \textbf{108}, no.8, 085010 (2023)
[arXiv:2211.11821 [gr-qc]].



\end{thebibliography}
\end{document}